\documentclass[10pt]{article}
\usepackage{amsmath}
\usepackage[mathcal]{euscript}
\usepackage{graphicx}
\def \myfigures #1#2#3#4#5#6#7#8
{\begin{figure}[ht]
    \begin{center}
        \includegraphics[width=#2 \textwidth]{#1.eps}
        \hfill
        \includegraphics[width=#6 \textwidth]{#5.eps}
        \parbox[t]{#4\textwidth}{\caption {#3}\label{#1}}
        \hfill
        \parbox[t]{#8\textwidth}{\caption {#7}\label{#5}}
    \end{center}
\end{figure} }

\begin{document}
\title{Dark energy model with variable $q$ and $\omega$ in LRS Bianchi-II space-time}

\author{Bijan Saha$^1$ and Anil Kumar Yadav$^{2\;\dag}$}

\footnotetext[1]{Laboratory of Information Technologies, Joint Institute for Nuclear Research
Dubna - 141 980, Russia.\;\; E-mail : bijan@jinr.ru}

\footnotetext[2]{Department of Physics, Anand Engineering
College, Keetham, Agra-282 007, India.\\
E-mail: abanilyadav@yahoo.co.in\\
$\dag$ corresponding author}

\date{}

\maketitle

\begin{abstract}
The present study deals with spatial homogeneous and
anisotropic locally rotationally symmetric (LRS)
Bianchi-II dark energy model in general relativity. The Einstein's
field equations have been solved exactly by taking into account the
proportionality relation between one of the components of shear
scalar $(\sigma^{1}_{1})$ and expansion scalar $(\vartheta)$, which,
for some suitable choices of problem parameters, yields time
dependent equation of state (EoS) and deceleration parameter (DP),
representing a model which generates a transition of universe from
early decelerating phase to present accelerating phase. The physical and
geometrical behavior of universe have been discussed in detail.\\
\end{abstract}



\section{Introduction}

The discovery of acceleration of the universe stands as a major
breakthrough of the observational cosmology. The power of
observations in cosmology is clear from the observations of
supernovae of Ia (SN Ia) which dramatically changed, about a decade
ago, the then standard picture of cosmology - of an expanding
universe evolving under the rules of general relativity such that
the expansion rate should slow down as cosmic time unfolds. Surveys
of cosmologically distant SN Ia
(Riess et al. 1998; Permutter et al. 1999) indicated the presence of new,
unaccounted - for $\it{dark \; energy}$ that opposes the self-attraction
of matter and causes the expansion of the universe to accelerate.
When combined with indirect measurements using cosmic microwave
background (CMB) anisotropies, cosmic shear and studies of galaxy
clusters, a cosmological world model has emerged that describes the
universe at flat, with about $70\%$ of it's energy contained in the
form of this cosmic dark energy (Seljak et al. 2005). This
acceleration is realized with negative pressure and positive energy
density that violate the strong energy condition. This violation
gives a reverse gravitational effect. Due to this effect, the
universe gets a jerk and the transition from the earlier
deceleration phase to the recent acceleration phase take place
(Caldwell et al 2002). The cause of this sudden transition and the source of
accelerated expansion is still unknown. In physical cosmology and
astronomy, the simplest candidate for the DE is the cosmological
constant $(\Lambda)$, but it needs to be extremely fine-tuned to
satisfy the current value of the DE density, which is a serious
problem. Alternatively, to explain the decay of the density, the
different forms of dynamically changing DE with an effective
equation of state (EoS), $\omega = p/\rho < -1/3$, were
proposed instead of the constant vacuum energy density. Other
possible forms of DE include quintessence $(\omega > - 1)$
(Steinhardt et al. 1999), phantom $(\omega < - 1)$ (Caldwell 2002) etc. While the the
possibility $\omega << - 1$ is ruled out by current cosmological
data from SN Ia (Supernovae Legacy Survey, Gold sample of Hubble
Space Telescope) (Riess et al. 2006; Astier et al.
2006), CMBR (WMAP, BOOMERANG) (Eisentein et al. 2005;
MacTavish et al. 2006) and large scale structure (Sloan
Digital Sky Survey) data (Komatsu et al. 2009), the dynamically evolving DE
crossing the phantom divide line (PDL) $(\omega = - 1)$ is mildly
favored. Some other limits obtained from observational results
coming from SNe Ia data (Knop et al 2003) and combination of SNe Ia data
with CMBR anisotropy and galaxy clustering statistics (Tegmark et al. 2004) are $-1.67 <
\omega < -0.62$ and $-1.33 < \omega < -0.79$, respectively. The
latest results in 2009, obtained after a combination of cosmological
data-sets coming from CMB anisotropies, luminosity distances of high
red-shift type Ia supernovae and galaxy clustering, constrain the
dark energy EoS to $-1.44 < \omega < -0.92$ at $68\%$
confidence level (Komatsu et al. 2009; Hinshaw et al. 2009)\\

Moreover, in recent years Bianchi universes have been gaining an
increasing interest of observational cosmology, since the WMAP data 
(Hinshaw et al. 2003, 2007; Jaffe et al. 2005) seem to require an addition to the
standard cosmological model with positive cosmological constant that
resembles the Bianchi morphology (Jaffe et al. 2006a, 2006b; Campanelli et al. 2006, 2007; Hoftuft et al. 2009).
According to this, the universe should achieve a slightly
anisotropic special geometry in spite of the inflation, contrary to
generic inflationary models and that might be indicating a
nontrivial isotropization history of universe due to the presence of
an anisotropic energy source. The Bianchi models isotropize at late times 
even for ordinary matter, and the possible anisotropy of the Bianchi metrics necessarily 
dies away during the inflationary era (Ellis 2006). In fact this isotropization 
of the Bianchi metrics is due to the implicit assumption that the DE is isotropic in nature. Therefore, 
the CMB anisotropy can also be fine tuned, since the Bianchi universe anisotropies 
determine the CMB anisotropies. The price of this property of DE is a voilation of null energy condition 
(NEC) since the DE crosses the phantom divide line (PDL), in particular depending on the direction. \\

The anomalies found in the cosmic microwave background (CMB) and
large scale structure observations stimulated a growing interest in
anisotropic cosmological model of universe. Here we confine
ourselves to models of Bianchi-type II. Bianchi type-II space-time
has a fundamental role in constructing cosmological models suitable
for describing the early stages of evolution of universe. Asseo and
Sol (1987) emphasized the importance of Bianchi type-II
Universe. Recently Pradhan et al (2011) and Kumar
and Akarsu (2011) have dealt with Bianchi-II DE models by
considering the spatial law of variation of Hubble's parameter which
yields the constant value of deceleration parameter (DP). Some
authors (Akarsu and Kilinc 2010a, 2010b; Yadav et al. 2011; Yadav and Yadav 2011; Kumar and Yadav
2011; Yadav 2011; Adhav et al. 2011 and
recently Yadav and Saha 2011) have studied DE models with
variable EoS parameter. In this paper, we presented general
relativistic cosmological model with time dependent DP in LRS
Bianchi-II space-time which can be described
by isotropic and variable EoS parameter. The paper is organized as
follows: The metric and field equation are presented in section 2.
Section 3 deals with the solution of field equations and physical
behavior of the model. Finally the findings of paper are discussed in section 4.   \\

\section{The metric and field equations}

The gravitational field in our case is given by a Bianchi type-II
(BII) metric
\begin{equation}
ds^{2} = - dt^2 + a_1^2(dx_1 - x_3 dx_2)^2 + a_2^2 dx_2^2 + a_3^2
dx_3^2, \label{BII}
\end{equation}
with $a_1,\,a_2,\,a_3$ being the functions of time only. In what
follows, we consider the LRS BII model setting $a_2 = a_3$.

Given the fact that the dark energy is isotropically distributed, it
is enough to consider only three Einstein equations (Saha 2011)
corresponding to the metric \eqref{BII}, namely
\begin{subequations}
\label{ein}
\begin{eqnarray}
2\frac{\ddot a_2}{a_2} + \Bigl(\frac{\dot a_2}{a_2}\Bigr)^2 -
\frac{3}{4} \frac{a_1^2}{a_2^4} &=&  - \omega \rho, \label{11}\\
\frac{\ddot a_1}{a_1} +\frac{\ddot a_2}{a_2} +\frac{\dot
a_1}{a_1}\frac{\dot a_2}{a_2} + \frac{1}{4}\frac{a_1^2}{a_2^4} &=&
- \omega \rho, \label{22} \\
2\frac{\dot a_1}{a_1}\frac{\dot a_2}{a_2} + \Bigl(\frac{\dot
a_2}{a_2}\Bigr)^2  - \frac{1}{4}\frac{a_1^2}{a_2^4} &=&  \rho.
\label{00}
\end{eqnarray}
\end{subequations}
Here over dots denote differentiation with respect to time ($t$).

Let us introduce a new function
\begin{equation}
V = a_1 a_2^2 = \sqrt{-g}. \label{V}
\end{equation}

The expressions for expansion and shear for BII metric given by
\eqref{BII} read:
\begin{equation}
\vartheta = u^\mu_{;\mu} = \Gamma^\mu_{\mu 0} =
\frac{\dot{a_1}}{a_1} + 2\frac{\dot{a_2}}{a_2} = \frac{\dot{V}}{V},
\label{expanBII}
\end{equation}
and
\begin{eqnarray}
\sigma_1^1 = \frac{\dot a_1}{a_1} - \frac{1}{3} \vartheta, \quad
\sigma_2^2 = \sigma_3^3 = \frac{\dot a_2}{a_2} - \frac{1}{3}
\vartheta, \quad \sigma_2^1 = z\Bigl[ \frac{\dot a_1}{a_1}
-\frac{\dot a_2}{a_2} + \frac{z^2 a_1 \dot a_1}{a_2^2} \Bigr].
\label{shearmix}
\end{eqnarray}

Let us now define the generalized and directional Hubble parameters.
As in know, the Hubble parameter was defined by E. Hubble for the
FRW model
\begin{equation}
ds^{2} = - dt^2 + a^2(dx_1^2 + dx_2^2 +  dx_3^2), \label{FRW}
\end{equation}
as $H = \frac{1}{a}\frac{da}{dt}$. Taking into account that
$\sqrt{-g} = a^3$ it can be defined as $H =
\frac{1}{3}\frac{1}{\sqrt{-g}}\frac{d \sqrt{-g}}{dt}$, and the
directional Hubble parameters as $H_{i} =
\frac{1}{\sqrt{g_{ii}}}\frac{d \sqrt{g_{ii}}}{dt}$ or $H_{i} =
\frac{1}{2}\frac{1}{g_{ii}}\frac{d g_{ii}}{dt}$.

Taking into account that for BII metric \eqref{BII} $\sqrt{-g} =
a_1a_2a_3 = a_1 a_2^2 = V$ and $g_{11} = a_1^2$, $g_{22} = x_3^2
a_1^2 + a_2^2$ and $g_{33} = a_3^2$,  analogically we define

\begin{equation}
H_{1} = \frac{\dot a_1}{a_1},\quad H_{2} = \frac{x_3^2 a_1\dot
a_1 + a_2 \dot a_2}{x_3^2a_1^2 + a_2^2},\quad H_{3} = \frac{\dot
a_2}{a_2}
\end{equation}
and
\begin{equation}
H =  \frac{1}{3}\frac{\dot V}{V} = \frac{1}{3}\bigl(\frac{\dot
a_1}{a_1} + 2 \frac{\dot a_2}{a_2}\bigr). \label{Hubble}
\end{equation}
It should be noted that though for $a_2 = a_1$ we have $H_{1} =
H_{2} = H_{3}$ as in isotropic case, the present definition does
not lead to $H = (H_{1} + H_{2} + H_{3})/3$. For this equality
to held, one must set $H_{2} = \frac{\dot a_2}{a_2}$.
Unfortunately, there is no unique definition for directional Hubble
parameters. Finally we define the deceleration parameter (DP) as
\begin{equation}
q = -\frac{V\ddot V}{\dot V^2}. \label{dec}
\end{equation}

Imposing the proportionality condition, i.e., assuming that the
expansion $\vartheta$ is proportional to say $\sigma_1^1$:
\begin{equation}
\vartheta \propto \sigma_1^1, \label{propcon}
\end{equation}
one finds the following relations between the metric functions
\begin{equation}
a_2 = a_1^n, \label{prop}
\end{equation}
with $n$ being some constant. Inserting \eqref{prop} into \eqref{V}
we obtain
\begin{equation}
a_1 = V^{1/(2n+1)}, \quad  a_2 = V^{n/(2n+1)}, \label{a12}
\end{equation}
Subtraction of \eqref{22} from \eqref{11} gives
\begin{equation}
\frac{\ddot a_2}{a_2} - \frac{\ddot a_1}{a_1} + \bigl(\frac{\dot
a_2}{a_2}\bigr)^2 - \frac{\dot a_1}{a_1}\frac{\dot a_2}{a_2} -
\frac{a_1^2}{a_2^4} = 0. \label{1122}
\end{equation}
Inserting $a_1$ and $a_2$ from \eqref{a12} into \eqref{1122} we find
equation for defining $V$:
\begin{equation}
\ddot V = \frac{2n+1}{n-1} V^{(3-2n)/(1+2n)}, \label{V2}
\end{equation}
with the solution in quadrature
\begin{equation}
\int \frac{dV}{\sqrt{V^{4/(2n+1)} + C}} =
\frac{2n+1}{\sqrt{2(n-1)}}\,t. \label{Vsol}
\end{equation}
Eq. \eqref{Vsol} imposes some restriction on the choice of $n$,
namely, $n > 1$. Thus we see that the proportionality condition
\eqref{propcon} in our case does not allow isotropization of the
initially anisotropic space-time.

Once $V$ is defined, we can define DP from \eqref{dec} and EoS
parameter from
\begin{eqnarray}
\omega &=& - \frac{4(n+1)(2n+1)V\ddot V - 4(n^2 + 2n){\dot V}^2 +
(2n+1)^2 V^{4/(2n+1)}}{4(n^2 + 2n){\dot V}^2 - (2n+1)^2
V^{4/(2n+1)}} \nonumber\\
&=& 1 - \frac{4(n+1)(2n+1)V\ddot V}{4(n^2 + 2n){\dot V}^2 - (2n+1)^2
V^{4/(2n+1)}}
\end{eqnarray}
Thus we see that $V$ plays central role here in defining all
physical quantities. In what follows we find $V$ from \eqref{V2} or
\eqref{Vsol} for some concrete values of $n$ or $C$.

\section{Solution of field equations}

One can not solve equation \eqref{Vsol} in general. So, in order to
solve the problem completely, we have to choose either $C$ or $n$ in
such a manner that equation \eqref{Vsol} be integrable. The easiest
way is to set $C= 0$ in \eqref{Vsol}. In that case one dully obtains
\begin{equation}
V = C_0 t^{(2n+1)/(2n-1)}, \quad C_0 =
\Bigl[\frac{(2n+1)^2}{(2n-1)\sqrt{2(n-1)}}\Bigr]^{(2n+1)/(2n-1)}.
\label{cq}
\end{equation}
As one sees, in this case $V$ is an increasing function of time, but
this solution leads to the constant DP.

Since, we are looking for a model explaining an expanding universe
with acceleration, we consider the case for nontrivial $C$, which
for a suitable choice of  $n$ gives the time dependent DP. The
motivation for time dependent DP is behind the fact that the
universe is accelerated expansion at present as observed in recent
observations of Type Ia supernova (Riess et al. 1998, 2004;
Perlmutter et al. 1999; Tonry et al. 2003;
Clocchiatti et al. 2006) and CMB anisotropies (Bennett et
al. 2003; de Bernardis et al. 2000; Hanany et al.
2000) and decelerated expansion in the past. Also, the
transition redshift from deceleration expansion to accelerated
expansion is about 0.5. Now for a Universe which was decelerating in
past and accelerating at the present time, the DP must show
signature flipping (see Padmanabhan and Roychowdhury 2003;
Amendola 2003; Riess et al. 2001). So, there is no
scope for a constant DP at present epoch. So, in general, the DP is
not a constant but
time variable.\\

Thus we consider the Eq. \eqref{Vsol} with a nontrivial $C$.  For $C
\ne 0$ Eq. \eqref{Vsol} allows exact solution only when $4/(2n+1) =
N$, where $N$ is an integer number. In this case $N$ can be integer
only for $n = 1/2$ and $n = 3/2$. Since $n > 1$ we have only one
option left, it is to choose $n = 3/2$. In this case Eq.
\eqref{Vsol} reduces to
\begin{equation}
\label{eq16}
\int\frac{dV}{\sqrt{V+C}}=4t
\end{equation}
which after integration leads
\begin{equation}
\label{eq17}
V=4t^{2}+2\beta t +\gamma
\end{equation}
where $\beta$ is the integrating constant and $\gamma = \frac{\beta^{2}}{4}-C$\\
Inserting equation \eqref{eq18} into \eqref{a12}, we obtain
\begin{equation}
\label{eq18}
a_{1}=(4t^{2}+2\beta t +\gamma)^{\frac{1}{4}}
\end{equation}
\begin{equation}
\label{eq19}
a_{2}=(4t^{2}+2\beta t +\gamma)^{\frac{3}{8}}
\end{equation}
The physical parameters such as directional Hubble's parameters
$(H_{{1}}, H_{{2}}, H_{{3}})$, average Hubble parameter $(H)$,
expansion scalar $(\theta)$ and scale factor $(a)$ are, respectively
given by
\begin{equation}
\label{eq20}
H_{{1}} = \frac{4t+\beta}{2(4t^{2}+2\beta t+\gamma)}
\end{equation}
\begin{equation}
 \label{eq21}
H_{{2}} = \frac{(4t+\beta)(x_{3}^{2}+12t^{2}+6\beta t+3\gamma)}{2(4t^{2}+2\beta t+\gamma)(x_{3}^{2}
+4t^{2}+2\beta t +\gamma)}
\end{equation}
\begin{equation}
\label{eq22}
H_{{3}}=\frac{3(4t+\beta)}{4(4t^{2}+2\beta t+\gamma)}
\end{equation}
\begin{equation}
 \label{eq23}
H = \frac{2(4t+\beta)}{3(4t^{2}+2\beta+\gamma)}
\end{equation}
\begin{equation}
 \label{eq24}
\theta = \frac{2(4t+\beta)}{(4t^{2}+2\beta+\gamma)}
\end{equation}
\begin{equation}
 \label{eq25}
a=(4t^{2}+2\beta t+\gamma)^{\frac{1}{3}}
\end{equation}
The components of shear scalar are given by
\begin{eqnarray}
 \label{eq26}
\sigma^{1}_{1}=-\frac{4t+\beta}{6(4t^{2}+2\beta t +\gamma)}, \quad\\
\sigma^{2}_{2}=\sigma^{3}_{3}=\frac{4t+\beta}{12(4t^{2}+2\beta t +\gamma)}, \quad\\
\sigma^{1}_{2}=\frac{x_{3}^{3}(4t+\beta)}{4(4t^{2}+2\beta t + \gamma)^{\frac{5}{4}}}-
\frac{x_{3}(4t+\beta)}{4(4t^{2}+2\beta t+\gamma)}
\end{eqnarray}
The value of DP $(q)$ is found to be
\begin{equation}
\label{eq29} q=-\frac{1}{2\left[1+\frac{C}{4t^{2}+2\beta t+\gamma}\right]}=-\frac{1}{2\left[1+\frac{C}{a^{3}}\right]}
\end{equation}
\begin{figure}
\begin{center}
\includegraphics[width=4.0in]{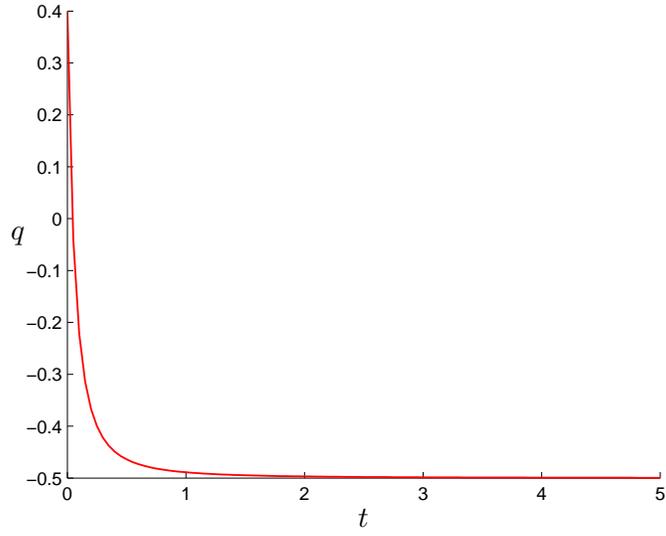}
\caption{Plot of deceleration parameter $(q)$ versus time $(t)$.}
\label{fg:abF1.eps}
\end{center}
\end{figure}
\begin{figure}
\begin{center}
\includegraphics[width=4.0in]{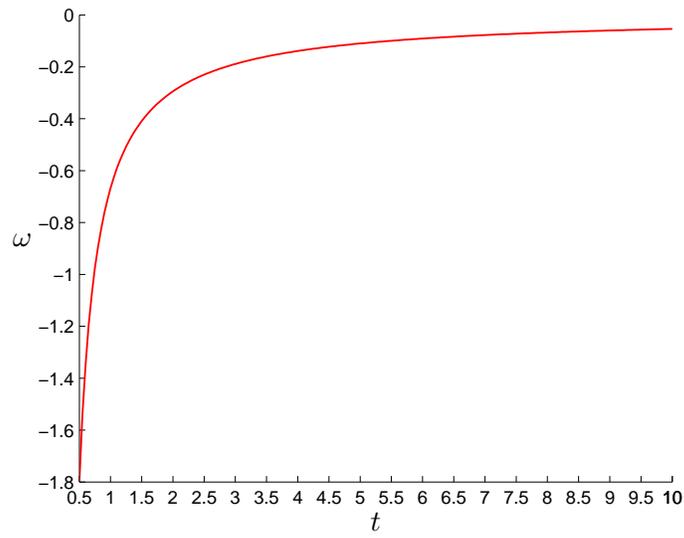}
\caption{Plot of EoS parameter $(\omega)$ versus time $(t)$.}
\label{fg:abF2.eps}
\end{center}
\end{figure}
The sign of $q$ indicates whether the model inflates or not. A
positive sign of $q$ corresponds to the standard decelerating model
whereas the negative sign of $q$ indicates indicates inflation. The
recent observations of SN Ia (Riess et al. 1998, Perlmutter et al. 1999) reveal
that the present universe is accelerating and the value of DP lies
somewhere in the range $-1 < q < 0$. Figure $1$ depicts the
variation of DP versus cosmic time as representative case with
appropriate choice of constants of
integration and other physical parameters.\\
The energy density of the cosmic fluid $(\rho)$, EoS parameter
$(\omega)$ and density parameter $(\Omega)$ are found to be
\begin{equation}
\label{eq30}
\rho = \frac{21(4t+\beta)^{2}}{16(4t^{2}+2\beta t +\gamma)^{2}} -
\frac{1}{4(4t^{2}+2\beta t +\gamma)}
\end{equation}
\begin{equation}
\label{eq31} \omega = 1 -\frac{320(4t^{2}+2\beta
t+\gamma)}{21(8t+2\beta)^{2}-16(4t^{2}+2\beta t+\gamma)}
\end{equation}
\begin{equation}
 \label{eq32}
\Omega = \frac{63}{64}-\frac{3(4t^{2}+2\beta t +\gamma)}{16(4t+\beta)^{2}}
\end{equation}

From Eq. \eqref{eq31} follows that at large $t$ when only the
quadratic terms stay alive, from EoS parameter we find
\begin{equation}
\label{dust} \omega \to 1 -\frac{320.4t^{2} }{21.(8t)^{2}-16.4t^{2}}
= 1 - \frac{320.4}{(21 - 1).64} = 0,
\end{equation}
i.e., under the present assumption the universe is ultimately filled
with dust only at remote future.\\

The initial time of the universe is
$t=\frac{-\beta+\sqrt{\beta^{2}-4\gamma}}{4}$. Therefore, at
$t=\frac{-\beta+\sqrt{\beta^{2}-4\gamma}}{4}$, the spatial volume
vanishes while all other parameter diverge. Thus the derived model
starts expanding with big bang singularity at
$t=\frac{-\beta+\sqrt{\beta^{2}-4\gamma}}{4}$ which can be shifted
to $t=0$ by choosing $\gamma = 0$. This singularity is point type
because the directional scale factors $a_{1}(t)$
and $a_{2}(t)$ vanish at initial moment. The components of shear scalar vanish at $t\rightarrow\infty$. Thus 
in derived model the initial anisotropy dies out at later time.\\

Figure 2 depicts the variation EoS parameter $(\omega)$ versus
cosmic time as representative case with appropriate choice of
constants of integration and other physical parameters. It is shown
that the growth of $\omega$ takes place with negative
sign. It should be emphasized that there is a number of models for dark
energy (quintessence, Chaplygin gas, phantom and many more) and
quest for the right one is still going on. The main idea for the
DE is a negative pressure, so one can try with a
negative EoS parameter. It should be noted that the quintessence is
given by a barotropic EoS only with negative parameter. We don't
call fluid a DE, we just construct DE in analogy with
fluid. Figure 3 demonstrates the behavior of density
parameter $(\Omega)$ versus cosmic time in the evolution of universe
as representative case with appropriate choice of constants of
integration and other physical parameters\\

\begin{figure}
\begin{center}
\includegraphics[width=4.0in]{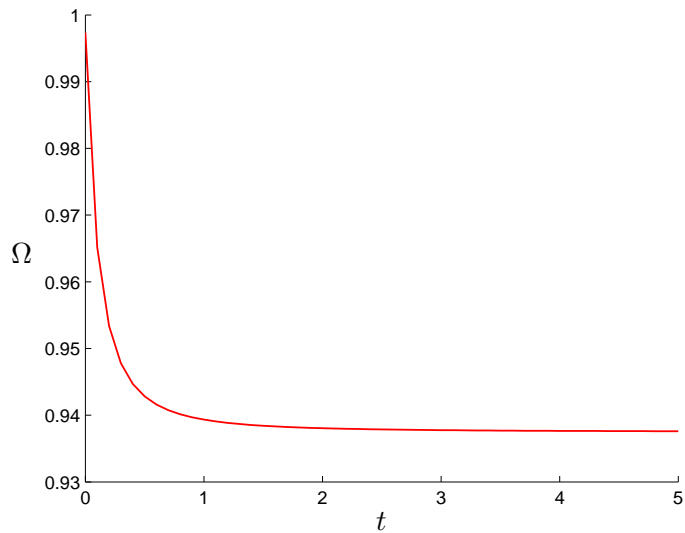}
\caption{Plot of density parameter $(\Omega)$ versus time $(t)$.}
\label{fg:abF4.eps}
\end{center}
\end{figure}

\section{Conclusion}
In this paper, we have investigated LRS Bianchi II DE model under
the assumption that $\vartheta \propto \sigma_{1}^{1}$. Under some
specific choice of problem parameters the present consideration
yields the variable DP and EoS parameter. It is to be noted that our
procedure of solving the field equations are altogether different
from what Pradhan et al (2011) have adapted. Pradhan et al (2011) 
have solved the field equations by considering the
variation law for generalized Hubble's parameter which gives the
constant value of DP and only the evolution takes place either in
accelerating or decelerating phase whereas we have considered the
proportionality condition $\vartheta \propto \sigma_{1}^{1}$ in such
a way that gives variable DP which evolves from decelerating phase
to current accelerating phase (see Fig. 1). Thus the present DE
model has transition of universe from the early deceleration phase
to current acceleration phase which is in good agreement with recent
observations (2006). The model has singular origin and the
universe is ultimately filled with dust only at remote future.\\

The theoretical arguments suggest and observational
data show, the universe was anisotropic at the early stage. Here we
are dealing not only with the present state of the universe, but
drawing a picture of the universe from the remote past to present
day. We use the Bianchi model as one of many models able to describe
initial anisotropy that dies away as the universe evolves. So though
the model is anisotropic in the past for small t but it becomes
isotropic as $t\rightarrow\infty$.
In the derived model, the EoS parameter $(\omega)$ is evolving with
negative sign which may be attributed to the current accelerated
expansion of universe. Hence from the theoretical perspective, the present model can 
be a viable model to explain the late time acceleration of the universe. In other words, 
the solution presented here can be one of the potential candidates to describe the present universe as well as 
the early universe.\\

\subsection*{Acknowledgments}
One of authors (AKY) is thankful to  The Institute of Mathematical
Science (IMSc), Chennai, India for providing facility and support
where part of this work was carried out. Bijan Saha is thankful to
joint Romanian-LIT, JINR, Dubna Research Project, theme no.
09-6-1060-2005/2013. Finally We would like to thank the anonymous referee for his
valuable questions which helped us to understand the depth of the
problem.



%

\begin{thebibliography}{100}
\bibitem{ref1}
Adhav, K. S., et al: Astrophys. Space Sc. {\bf 331}, 689 (2011) 
\bibitem{ref3}
Akarsu, \"{O.}, and Kilinc, C. B.: Gen. Relativ. Gravit. \textbf{42}, 119 (2010a)
\bibitem{ref4}
Akarsu, \"{O.}, and Kilinc, C. B.: Gen. Relativ. Gravit. \textbf{42}, 763 (2010b)
\bibitem{ref5}
Amendola, L.: Mon. Not. R. Astron. Soc. \textbf{342}, 221 (2003) 
\bibitem{ref6}
Asseo, E., and Sol, H.: Phys. Rev. D \textbf{148}, 307 (1987) 
\bibitem {ref7}
Astier, P., et al: Astron. Astrophys. \textbf{447}, 31 (2006) 
\bibitem{ref8}
Bennett, C. L., et al: Astrophys. J. Suppl. Ser. \textbf{148}, 1 (2003) 
\bibitem{ref9}
Caldwell, R. R., et al: Phys. Rev. D \textbf{73}, 023513 (2006)
\bibitem {ref10}
Caldwell, R. R.: Phys. Lett. B \textbf{545}, 23 (2002) 
\bibitem{ref11}
Campanelli, L., Cea. P. and Tedesco, L.: Phys. Rev. D \textbf{97}, 131302 (2006) 
\bibitem{ref12}
Campanelli, L., Cea, P. and Tedesco, L.: Phys. Rev. D \textbf{76}, 063007 (2007)
\bibitem{ref13}
Clocchiatti, A., et al: Astrophys. J. \textbf{642}, 1 (2006) 
\bibitem{ref15}
de-Bernadis, P., et al: Nature \textbf{404}, 955 (2000) 
\bibitem {ref16}
Eisentein, D. J., et al: Astrophys. J. \textbf{633}, 560 (2005)
\bibitem{ref16a}
Ellis, G. F. R.: Gen. Relat. Gravit. \textbf{38}, 1003 (2006)
\bibitem{ref17}
Hanany, S., et al: Astrophys. J. \textbf{545}, L5 (2000)
\bibitem{ref18}
Hinshaw et al.: Astrophys. J. Suppl. \textbf{148}, 135 (2003)
\bibitem{ref19}
Hinshaw, G., et al: Astrophys. J. Suppl. \textbf{180}, 225 (2009) 
\bibitem{ref20}
Hinshaw, G., et al: Astrophys. J. Suppl. \textbf{148}, 135 (2009) 
\bibitem{ref21}
Hinshaw, G., et al: Astrophys. J. Suppl. \textbf{170}, 288 (2007)
\bibitem{ref22}
Hoftuft, J., et al: Astrophys. J. \textbf{699}, 985 (2009)  
\bibitem{ref23}
Jaffe, J., et al: Astrophys. J. \textbf{629}, L1 (2005) 
\bibitem{ref24}
Jaffe, J., et al: Astrophys. J. \textbf{643}, 616 (2006a) 
\bibitem{ref25}
Jaffe, J., et al: Astron. Astrophys. \textbf{460}, 393 (2006b) 
\bibitem{ref26}
Knop, R. K., et al: Astrophys. J. \textbf{598}, 102 (2003) 
\bibitem {ref27}
Komatsu, E., et al: Astrophys. J. Suppl. Ser. \textbf{180}, 330 (2009) 
\bibitem{ref28}
Kumar, S. and Akarsu, \"{O.}: arXiv: 1110.2408 [gr-qc] (2011)
\bibitem{ref29}
Kumar, S. and Yadav, A. K.: Mod. Phys. Lett. A {\bf 26}, 647 (2011) 
\bibitem {ref30}
MacTavish, C. J., et al: Astrophys. J. \textbf{647}, 799 (2006) 
\bibitem{ref31}
Padmanabhan, T. and Raychowdhury, T.: Mon. Not. R. Astron. Soc. \textbf{344}, 823 (2003) 
\bibitem{ref32}
Permutter, S., et al: Astrophys. J. \textbf{517}, 565 (1999)
\bibitem{ref33}
Pradhan, A., Amirhashchi, H. and Jaiswal, R.: Astrophys. Space Sc. \textbf{334}, 249 (2011)
\bibitem {ref34}
Riess, A. G., et al: Astron. J. \textbf{116}, 1009 (1998) 
\bibitem {ref35}
Riess, A. G., et al: Astron. J. \textbf{607}, 665 (2004)
\bibitem{ref36}
Riess, A. G., et al: Astrophys. J. \textbf{560}, 49 (2001)
\bibitem{ref37}
Saha, B.: Cent. European J. Phys. {\bf 9}, 939 (2011) 
\bibitem{ref38}
Seljak, et al: Phys. Rev. D \textbf{71}, 043511 (2005) 
\bibitem{ref39}
Steinhardt, P. J., Wang, L. M. and Zlatev, I.: Phys. Rev. D {\bf59}, 123504 ( 1999) 
\bibitem{ref40}
Tegmark, M., et al: Phys. Rev. D \textbf{69}, 103501 (2004)
\bibitem{ref41}
Tonry, J. L., et al: Astrophys. J. \textbf{594}, 1 (2003)  
\bibitem{ref42}
Yadav, A. K. and Yadav, L.: Int. J. Theor. Phys. {\bf 50}, 218 (2011) 
\bibitem {ref43}
Yadav, A. K., Rahaman, F. and Ray, S.: Int. J. Theor. Phys. {\bf 50},
871 (2011)
\bibitem{ref44}
Yadav, A. K.: Astrophys. Space Sc. \textbf{335}, 565 (2011)
\bibitem{ref45}
Yadav, A. K. and Saha, B.: Astrophys. Space Sc. \textbf{337}, 759 (2012)










\end{thebibliography}
\end{document}